\documentstyle[aps]{revtex}
\newcommand{\bb}{\begin{equation}}
\newcommand{\ee}{\end{equation}}
\newcommand{\ba}{\begin{array}}
\newcommand{\ea}{\end{array}}

\begin {document}

\title{Quantum-mechanical model for particles carrying
        electric charge and magnetic flux in two dimensions
        \thanks{published in Phys. Rev. A {\bf 59} (1999) 3228-3235.
        \copyright The American Physical Society}}
\author{Qiong-gui Lin\thanks{E-mail addresses: qg\_lin@163.net,
        stdp@zsu.edu.cn}}
\address{China Center of Advanced Science and Technology (World
	Laboratory),\\
        P.O.Box 8730, Beijing 100080, People's Republic of China\\
        and\\
        Department of Physics, Zhongshan University, Guangzhou
        510275,\\
        People's  Republic of China \thanks{Mailing address}}

\maketitle

\begin{abstract}
{\normalsize
We propose a simple quantum mechanical equation for $n$ particles
in two dimensions, each particle carrying electric charge and
magnetic flux. Such particles appear in (2+1)-dimensional
Chern-Simons field theories as charged vortex soliton solutions,
where the ratio of charge to flux is a constant independent of the
specific solution. As an approximation, the charge-flux interaction
is described here by the Aharonov-Bohm potential, and the
charge-charge interaction by the Coulomb one. The equation for two
particles, one with charge and flux ($q$, $\Phi/Z$) and the other
with ($-Zq$, $-\Phi$) where $Z$ is a pure number is studied in
detail. The bound state problem is solved exactly for arbitrary $q$
and $\Phi$ when $Z>0$. The scattering problem is exactly solved
in parabolic coordinates in special cases when $q\Phi/2\pi\hbar c$
takes integers or half integers. In both cases the cross sections
obtained are rather different from that for pure Coulomb scattering.}
\end{abstract}

\leftline {PACS number(s): 03.65.Bz, 12.90.+b}

\section{Introduction}               

Field theories with Chern-Simons (CS) term in (2+1)-dimensional
space-time admit soliton solutions carrying both electric charge
and magnetic flux [1-8]. These solutions are often called CS
vortices or vortex solitons, as compared with Nielsen-Olesen
vortices [9] which are electrically neutral. They appear in both
relativistic and nonrelativistic field theories, and regardless of
whether the gauge field action involves both Maxwell and CS terms
or only a pure CS term. The ratio of electric charge $q$ to magnetic
flux $\Phi$ depends only on the parameters in the field theoretical
model, not on the specific solution. Such solutions are not only
of interest in field theories, but also expected to be useful in
condensed matter physics. However, the interaction of these vortex
solitons is very complicated. A single soliton solution is available
in analytic form only for nonrelativistic theory and when the Maxwell
term is absent. It seems difficult to find multi-soliton solutions
in closed forms, especially when both Maxwell and CS terms are
present. Therefore a simple quantum mechanical model for the
interaction of such vortex solitons may be of interest. The purpose
of the present paper is to study such a model.

The real CS vortices have finite sizes. The electric charge density
and the magnetic flux density (the magnetic field) depend on the
specific solution. As a simple approximation, we use point-like
particles to represent them in this paper. Both the magnetic flux
and the electric charge are then confined to a region of infinitesimal
area, in other words, to a point where the particle is located. The
vector potential associated with the flux is the Aharonov-Bohm (AB)
potential [10]. (see also Refs. [11,12] for some more works on the
subject.)  This is responsible to the
charge-flux interaction. As for the charge-charge interaction, we make
use of the Coulomb potential. Note that in two-dimensional space there
are two kinds of Coulomb potentials. The first one satisfies the
two-dimensional Poisson equation with point source and is proportional
to $\ln r$, where $r$ is the distance between the two point charges.
The second simply imitates the form of the three-dimensional one and
is proportional to $1/r$. It should be remarked that the real
interaction between the CS vortices may be very complicated, and it
depends on whether the field theoretical model involves both Maxwell
and CS terms or only a CS term. Neither of the above forms can be
expected to be capable of well describing the real situation. Either
one is in any case a rough approximation. We prefer the latter one
since it is easier to obtain exact solutions in this case. This is
the potential adopted in the study of the so called two-dimensional
hydrogen atom (2H) [13-18].

In this paper we confine ourselves to the framework of nonrelativistic
quantum mechanics. Now that the forms of the interaction potentials
are established, we can write down an $n$-body Schr\"odinger equation
for these particles carrying magnetic flux
as well as electric charges.
This is done in Sec. II. The $a$th particle has charge and flux
($q_a$, $\Phi_a$), where $a=1,2,\ldots,n$. It should be emphasized
that the ratio  $q_a/\Phi_a$ does not depend on $a$, as pointed out
in the first paragraph. After the time variable is separated out to
obtain a stationary Schr\"odinger equation, we concentrate our
attention on the two-body problem. This is separable into two
equations. One governs the center-of-mass motion, which is free, and
the other governs the relative motion, which is of main interest to
us and is the main subject of the remaining part of this paper. It
is remarkable that the separability of the two-body equation crucially
depends on the condition $q_1/\Phi_1=q_2/\Phi_2$. We then denote
$(q_1, \Phi_1)=(q, \Phi/Z)$, $(q_2, \Phi_2)=(-Zq, -\Phi)$, where
$Z$ is a nonvanishing real number. The relative Hamiltonian has the
same form as that for a particle of reduced mass moving in the
composite field of a vector AB potential and a scalar Coulomb one.
This may be called an Aharonov-Bohm-Coulomb (ABC) system. Although
the so called ABC system has been dealt with by numerous works
[19-24] in the literature, the Coulomb potential considered there is
a three-dimensional one. Thus the situation is quite different from
that studied here. In other words, the model studied in the above
cited works is a three-dimensional ABC system, while that encountered
here is a two-dimensional one.

In Sec. III we study the bound state problem. Bound states are
possible only when $Z>0$, i.e., when the Coulomb field represents
attractive force, regardless of whether an AB potential is present.
When $\Phi=0$, the spectrum is just those of the 2H. The level $E_N$
has degeneracy $2N+1$ ($N=0,1,2,\ldots$). If $q\Phi/2\pi\hbar c$
takes nonvanishing integers, the spectrum is roughly the same except
that the ground state has energy $E_1$ and the level $E_N$ has
degeneracy $2N$ ($N=1,2,\ldots$) since some solutions are not
acceptable. In the general case each level $E_N$ of the 2H splits
into two, each with lower degeneracy. When $q\Phi/2\pi\hbar c$ takes
half integers, however, some of the splited levels coincide and we
have again a high degeneracy. The degeneracy implies that the system
should  have SU(2) symmetry in this case, as the SO(3) symmetry of
the ordinary 2H [13, 16-17]. But this has not been explicitly proved.

In Sec. IV we study the scattering problem. In the general case
partial wave expansion in the polar coordinates should be employed.
However, as the asymptotic form of the partial wave involves
logarithmic distortion due to the long range nature of the Coulomb
field, it is somewhat difficult to handle the partial wave expansion.
In this paper we restrict our discussion to special cases where
$q\Phi/2\pi\hbar c$ takes integers or half integers. In these cases
the scattering problem can be exactly solved in parabolic coordinates,
as the ordinary Coulomb scattering in two dimensions [25]. Note that
what we use here are parabolic coordinates on the plane, and thus they
are quite different from the rotational parabolic coordinates used in
the discussion of the ordinary three-dimensional
Coulomb problem in the text books of quantum mechanics.
The latter are also used in the study of the
three-dimensional ABC system [19]. When $\Phi=0$ the cross
section is just that for the Coulomb scattering in two dimensions.
When $q\Phi/2\pi\hbar c$ takes nonzero integers, the cross section
gains an additional term which comes from the interference of the
scattered wave with an additional stationary wave present in the
scattering solution. Without the stationary wave term the solution
would become meaningless at the origin.  To the best of our knowledge,
such circumstances are not encountered previously in the literature.
When $q\Phi/2\pi\hbar c$ takes half integers, the result is simple
but of course rather different from that for pure Coulomb scattering.
Without the Coulomb field our results reduce to those for pure AB
scattering [10-11].
The classical limit of the results is also discussed.

Sec. V is devoted to a brief summary and some more remarks.

\section{The model}          

Consider $n$ point-like particles carrying magnetic flux as well as
electric charges in two-dimensional space. The $a$th particle has
mass $\mu_a$, carries electric charge and magnetic flux
($q_a$, $\Phi_a$), $a=1,2,\ldots,n$. The position of the $a$th
particle is denoted by ${\bf r}_a=(x_a, y_a)$. As remarked in the
introduction, the ratio $q_a/\Phi_a$ is independent of $a$. More
precisely, we have
\bb
{q_1\over\Phi_1}={q_2\over\Phi_2}=\ldots={q_n\over\Phi_n}.
\ee        
We describe the charge-flux interactions among the particles by the
vector AB potentials and the charge-charge interactions by the scalar
Coulomb ones. The $n$-body wave function is denoted by
$\Psi^{(n)}(t, {\bf r}_1,\ldots,{\bf r}_n)$. In this paper we work in
the domain of nonrelativistic quantum mechanics. The Schr\"odinger
equation for the wave function is then
$$
i\hbar{\partial\Psi^{(n)}\over\partial t}=H_{\rm T}\Psi^{(n)},
\eqno(2{\rm a})
$$
where $H_{\rm T}$ is the Hamiltonian of the system given by
$$
H_{\rm T}=-\sum_{a=1}^n{\hbar^2\over 2\mu_a}\left[\nabla_a
-{iq_a\over\hbar c}{\bf A}_a({\bf r}_1,\ldots,{\bf r}_n)\right]^2
+\sum_{a<b}{q_a q_b\over|{\bf r}_a-{\bf r}_b|},
\eqno(2{\rm b})
$$
where the second term (the Coulomb interaction) involves a double
summation subject to the condition $a<b$, and
${\bf A}_a({\bf r}_1,\ldots,{\bf r}_n)$ is the AB vector potential
at the position ${\bf r}_a$. Note that all particles, except the
$a$th one, contribute to ${\bf A}_a$. Thus the components of
${\bf A}_a$ are given by
$$
A_{ax}({\bf r}_1,\ldots,{\bf r}_n)=-\sum_{b\ne a}{\Phi_b\over 2\pi}
{y_a-y_b\over|{\bf r}_a-{\bf r}_b|^2},
$$
$$
A_{ay}({\bf r}_1,\ldots,{\bf r}_n)=\sum_{b\ne a}{\Phi_b\over 2\pi}
{x_a-x_b\over|{\bf r}_a-{\bf r}_b|^2}.
\eqno(2{\rm c})
$$
Since the Hamiltonian $H_{\rm T}$ does not involve $t$,
the time dependent
factor in $\Psi^{(n)}$ can be separated out. Let
\addtocounter{equation}{1}
\bb
\Psi^{(n)}(t, {\bf r}_1,\ldots,{\bf r}_n)=e^{-iE_{\rm T} t/\hbar}
\psi^{(n)}(t, {\bf r}_1,\ldots,{\bf r}_n),
\ee       
we have for $\psi^{(n)}$ the stationary Schr\"odinger equation
\bb
H_{\rm T}\psi^{(n)}=E_{\rm T}\psi^{(n)}.
\ee     

In the following we concentrate our attention on the two-body
problem, since this is the only case where exact analysis is possible.
In this case the first summation in Eq. (2b) contains two terms
and the second contains only one. We introduce the relative position
{\bf r} and the center-of-mass position {\bf R} defined by
\bb
{\bf r}={\bf r}_1-{\bf r}_2,\quad
{\bf R}={\mu_1{\bf r}_1+\mu_2{\bf r}_2\over M},
\ee     
Where $M=\mu_1+\mu_2$ is the total mass of the system. Note that
both ${\bf A}_1$ and ${\bf A}_2$ depend only on {\bf r}, it is not
difficult to recast $H_{\rm T}$ in the form
\begin{eqnarray}
H_{\rm T}&=&-{\hbar^2\over 2\mu_1}\left(\nabla_r-{iq_1\over \hbar c}
{\bf A}_1\right)^2-{\hbar^2\over 2\mu_2}\left(\nabla_r+{iq_2\over
\hbar c}{\bf A}_2\right)^2+{q_1 q_2\over r}\nonumber\\
&&-{\hbar^2\over 2M}\nabla_R^2+{i\hbar\over Mc}(q_1{\bf A}_1+
q_2{\bf A}_2)\cdot\nabla_R,
\end{eqnarray}          
where $r=|{\bf r}|$. Using Eqs. (1) and (2c), it can be shown that
\bb
q_1{\bf A}_1=-q_2{\bf A}_2.
\ee     
Thus Eq. (6) reduces to
\bb
H_{\rm T}=-{\hbar^2\over 2\mu}\left(\nabla_r-{iq_1\over \hbar c}
{\bf A}_1\right)^2+{q_1 q_2\over r}-{\hbar^2\over 2M}\nabla_R^2,
\ee     
where $\mu=\mu_1\mu_2/(\mu_1+\mu_2)$ is the reduced mass of the
system. Now Eq. (4) can be separated into two equations. Let
\bb
\psi^{(2)}({\bf r}_1,{\bf r}_2)=\psi_{\rm cm}({\bf R})\psi({\bf r}),
\ee     
we have
\bb
-{\hbar^2\over 2M}\nabla_R^2\psi_{\rm cm}=E_{\rm cm}\psi_{\rm cm},
\ee     
$$
H\psi=E\psi,\eqno(11{\rm a})
$$
where
$$
H=-{\hbar^2\over 2\mu}\left(\nabla_r-{iq_1\over \hbar c}
{\bf A}_1\right)^2+{q_1 q_2\over r},\eqno(11{\rm b})
$$
$$
A_{1x}=-{\Phi_2\over 2\pi}{y\over r^2},\quad
A_{1y}={\Phi_2\over 2\pi}{x\over r^2},\eqno(11{\rm c})
$$
and $E_{\rm cm}+E=E_{\rm T}$. Eq. (10) governs the center-of-mass
motion of the system, which is obviously free and will not be
discussed any further. Eq. (11) governs the relative motion of the
two particles, which is of essential interest to us and is the main
subject of the remaining part of this paper. In the following we omit
the subscript $r$ of $\nabla_r$. We also denote
$(q_1,\Phi_1)=(q,\Phi/Z)$ and $(q_2,\Phi_2)=(-Zq,-\Phi)$, where $Z$
is a nonvanishing real number. The Hamiltonian (11b) can be written
as
\addtocounter{equation}{1}
\bb
H=-{\hbar^2\over 2\mu}\left(\nabla+i{q\Phi\over 2\pi\hbar c}
\nabla\theta\right)^2-{Zq^2\over r},
\ee     
where $(r,\theta)$ are polar coordinates on the $xy$ plane, and $r$
has been used above.

As pointed out in the introduction, the Hamiltonian (12) is the same
as that governs the motion of a charged particle in the combined
field of a vector AB potential and a scalar Coulomb one. However, it
is quite different from that for the so called ABC system studied in
the literature, since that is a three-dimensional model while ours is
a two-dimensional one. More precisely, in their Coulomb potential,
$r=(x^2+y^2+z^2)^{1/2}$, whereas in ours $r=(x^2+y^2)^{1/2}$. In fact,
everything is independent of $z$ here, or, if one prefers, there is no
$z$ component here.

To conclude this section we emphasize that the separability of Eq. (4)
(for $n=2$) crucially depends on the relation (7) and thus on the
condition (1).

\section{Bound states}        

In this section we study bound states of the two-body system. These
are solutions vanishing at infinity of Eq. (11). It is convenient to
solve Eq. (11a), with the Hamiltonian written in the form of Eq. (12),
in polar coordinates. We denote
\bb
{q\Phi\over 2\pi\hbar c}=m_0+\nu,
\ee     
where $m_0$ is an integer and $0\le\nu<1$. Eq. (11) can be written in
polar coordinates as
\bb
{1\over r}{\partial\over\partial r}\left(r{\partial\psi\over
\partial r}\right)+{1\over r^2}\left({\partial\over\partial\theta}
+im_0+i\nu\right)^2\psi+\left({2\mu E\over\hbar^2}+{2\mu Zq^2\over
\hbar^2 r}\right)\psi=0.
\ee     
We write $\psi$ as
\bb
\psi(r,\theta)=R(r)e^{i(m-m_0)\theta},\quad m=0,\pm1,\pm2,\ldots,
\ee     
then $R(r)$ satisfies the equation
\bb
{d^2R\over dr^2}+{1\over r}{dR\over dr}
+\left[{2\mu E\over\hbar^2}+{2\mu Zq^2\over\hbar^2 r}-
{(m+\nu)^2\over r^2}\right]R=0.
\ee     
Now it can be shown that $E>0$ gives scattering solutions which will
not be discussed in this section. Thus bound states have $E<0$. It
will also become clear in the following that bound states are possible
only when $Z>0$, i.e., when the Coulomb potential represents
attraction. These are all familiar conclusions in the pure Coulomb
problem in three or two dimensions. Note that the factorized form of
the solution (15) itself requires
\bb
R(0)=0
\ee     
except for $m=m_0$. This is because $\theta$ is not well defined at
the origin. It will exclude some well behaved solutions of Eq. (16).
As $E<0$, we introduce a dimensionless variable $\rho$ defined as
\bb
\rho=\alpha r,\quad \alpha={\sqrt{-8\mu E}\over\hbar},
\ee     
and a new parameter
\bb
\lambda={Zq^2\over\hbar}\sqrt{-{\mu\over 2E}},
\ee     
then Eq. (16) can be written as
\bb
{d^2R\over d\rho^2}+{1\over \rho}{dR\over d\rho}
+\left[-{1\over 4}+{\lambda\over\rho}-
{(m+\nu)^2\over \rho^2}\right]R=0.
\ee     
Now we define a new function $u(\rho)$ through the relation
\bb
R(\rho)=e^{-\rho/2}\rho^{|m+\nu|}u(\rho),
\ee     
then we have for $u(\rho)$ the equation
\bb
\rho{d^2u\over d\rho^2}+(2|m+\nu|+1-\rho){du\over d\rho}
-\left(|m+\nu|+{1\over 2}-\lambda\right)u=0.
\ee     
This is the confluent hypergeometric equation. It is solved by the
confluent hypergeometric function
\bb
u(\rho)=CF(|m+\nu|+\textstyle{1\over 2}-\lambda,
2|m+\nu|+1,\rho),
\ee             
where $C$ is a normalization constant to be determined below. The
other solution to Eq. (22) makes $R(r)$ infinite at $r=0$ and is
thus dropped. The above solution, though well behaved at $r=0$,
diverges when $r\to\infty$: $u(\rho)$ behaves like $e^\rho$ and
$R(\rho)$ like $e^{\rho/2}$. Therefore it is still not acceptable in
general. Physically acceptable solutions appear when $E$ or $\lambda$
takes special values so that the confluent hypergeometric series
terminates. This happens when
\bb
|m+\nu|+\textstyle{1\over 2}-\lambda=-n_r, \quad
n_r=0,1,2,\ldots,
\ee     
and $u(\rho)$ becomes a polynomial of order $n_r$. From Eq. (19) we
see that this can be satisfied only when $Z>0$, and the energy
levels are given by
\bb
E=-{\mu Z^2 q^4\over 2\hbar^2(n_r+|m+\nu|+1/2)^2}.
\ee     
The corresponding wave function is
\bb
\psi_{n_r m}(r,\theta)=C_{n_r m}e^{-\rho/2}\rho^{|m+\nu|}
F\left(-n_r, 2|m+\nu|+1,\rho\right)e^{i(m-m_0)\theta}.
\ee     
There are degeneracies in the energy levels. This is why we have not
attached any subscript to $E$. The degeneracy depends on the values
of $\nu$ and $m_0$. The various cases are discussed as follows.

1. $\nu=m_0=0$. This is the case of a pure Coulomb problem, or the
2H. We introduce the principal quantum number
\bb
N=n_r+|m|,
\ee     
then the energy levels are written as
\bb
E_N=-{\mu Z^2 q^4\over 2\hbar^2(N+1/2)^2},\quad N=0,1,2,\ldots.
\ee     
With a given $N$, the possible values for ($n_r,m$) are ($N,0$),
($N-1,\pm1$), $\ldots$, ($0,\pm N$), and the degeneracy is
$d_N=2N+1$. These results are well known [13-18].

2. $\nu=0$, $m_0\ne0$. In other words, $q\Phi/2\pi\hbar c$ takes
nonzero integers. In this case the energy levels are roughly the same.
However, from  Eq. (26) we see that the solution with $m=0$ is not
acceptable, regardless of the value of $n_r$, because the radial part
of the wavefunction does not satisfy Eq. (17). Therefore the ground
state has energy $E_1$, and the level $E_N$ has degeneracy $d_N=2N$
($N=1,2,\ldots$).

3. $0<\nu<{1\over 2}$ or ${1\over 2}<\nu<1$. In this case each
level of the 2H
splits into two. When $m\ge 0$ we have
$$
E_N^+=-{\mu Z^2 q^4\over 2\hbar^2(N+\nu+1/2)^2},\quad N=0,1,2,\ldots,
\eqno(29{\rm a})
$$
while when $m<0$ we have
$$
E_N^-=-{\mu Z^2 q^4\over 2\hbar^2(N-\nu+1/2)^2},\quad N=1,2,\ldots.
\eqno(29{\rm b})
$$
The possible values of ($n_r,m$) correspond to $E_N^+$ are ($N,0$),
($N-1,1$), $\ldots$, ($0,N$), thus the degeneracy is $d_N^+=N+1$.
Those correspond to $E_N^-$ are ($N-1,-1$),
($N-2,-2$), $\ldots$, ($0,-N$), thus the degeneracy is $d_N^-=N$.
The difference between the case $0<\nu<{1\over 2}$ and the case
${1\over 2}<\nu<1$
lies in the order of the energy levels.
In the first case the order of the levels is
$$
E_0^+<E_1^-<E_1^+<\ldots<E_N^-<E_N^+<E_{N+1}^-<\ldots.
\eqno(30{\rm a})
$$
In the second case it is
$$
E_1^-<E_0^+<E_2^-<\ldots<E_N^-<E_{N-1}^+<E_{N+1}^-<\ldots.
\eqno(30{\rm b})
$$
\addtocounter{equation}{2}

4. $\nu={1\over 2}$. In other words, $q\Phi/2\pi\hbar c$ takes half
integers. In this case we have
\bb
E_N^+=E_{N+1}^-=-{\mu Z^2 q^4\over 2\hbar^2(N+1)^2},\quad N=0,1,2,
\ldots.
\ee          
The degeneracy of the level is $d_N^++d_{N+1}^-=2N+2$. This implies
that the system has higer dynamical symmetry than the geometrical
SO(2). It is well know that the 2H possesses SO(3) symmetry, just like
the ordinary three-dimensional hydrogen atom possesses  SO(4)
symmetry. It seems that the symmetry for the present case is SU(2),
and the above energy level corresponds to the value
$(N+{1\over 2})(N+{3\over 2})$
for the Casimir operator of the SU(2) algebra. But this has not been
explicitly proved. One can construct the Runge-Lenz vector in a way
similar to that in the case of 2H [16]. However, the conservation of
it and the closure of the algebra involve some difficulty due to the
singularity of the AB potential at the origin. Perhaps some other
method should be employed to deal with the problem.

Both bound state and scattering problems of the two-dimensional
Coulomb field can be solved in parabolic coordinates [17-18, 25].
Here we point out that the case 2. and 4. disscussed above can also
be solved in parabolic coordinates. As no new result can be obtained,
we will not discuss the solutions in detail. In the next section we
will deal with the scattering problem. It  is in these two cases that
exact solutions are available.

Finally we give the value of the normalization constant $C_{n_r m}$ in
the wave function (26):
\bb
C_{n_r m}={4\mu Zq^2\over \hbar^2(2n_r+2|m+\nu|+1)\Gamma(2|m+\nu|+1)}
\left[{\Gamma(n_r+2|m+\nu|+1)\over 2\pi n_r!(2n_r+2|m+\nu|+1)}\right]
^{1\over 2}.
\ee              

\section{Scattering problem}           
In this section we study scattering problem of the two-body system.
Here the Coulomb field may be either attractive or repulsive.
We denote $\kappa=Zq^2$, which may be positive or negative. For
general value of $m_0$ and $\nu$, one may employ the method of partial
wave expansion in polar coordinates. Then the starting point may be
Eqs. (15) and (16). However, the asymptotic form of $R(r)$ when $r\to
\infty$ involves the $\ln r$ distortion, due to the long-range nature
of the Coulomb field. This may be more clearly seen in the following.
Thus it is not easy to treat the partial wave expansion and to obtain
the scattering cross section in a closed form. For this reason we
confine ourselves in this paper to two special cases where exact
analysis can be carried out in parabolic coordinates, and defer the
general discussion to subsequent study.

Consider Eqs. (11a) and (12). Let us make a transformation
\bb
\psi(r,\theta)=e^{-i(m_0+\nu)\theta}\psi_0(r,\theta).
\ee     
The new wavefunction $\psi_0(r,\theta)$ satisfies the Schr\"odinger
equation with a pure Coulomb field:
\bb
-{\hbar^2\over 2\mu}\nabla^2\psi_0-{\kappa\over r}\psi_0=E\psi_0.
\ee          
In parabolic coordinates this equation can be  separated into two
ordinary differential equations while Eq. (11) cannot be separated.
The probability current density
\bb
{\bf j}={\hbar\over 2i\mu}(\psi^*\nabla\psi-\psi\nabla\psi^*)+
{(m_0+\nu)\hbar\over \mu}\psi^*\psi\nabla\theta
\ee           
can be written in terms of $\psi_0$ as
\bb
{\bf j}={\hbar\over 2i\mu}(\psi_0^*\nabla\psi_0-\psi_0\nabla\psi_0^*).
\ee           
Although $\psi_0$ satisfies a simpler equation, the problem does not
become easier since $\psi_0$ must satisfy a nontrivial boundary
condition
\bb
\psi_0(r,\theta+2\pi)=e^{i2\pi\nu}\psi_0(r,\theta)
\ee     
such that $\psi(r,\theta)$ is single valued. Moreover,
$\psi_0(r,\theta)$ should have proper behavior at the origin, so that
$\psi$ is well defined there. The latter condition also imposes a
constraint on the solution.

It is in general difficult to deal with Eq. (37). In the following we
only consider two special cases. The first is $\nu=0$, or $q\Phi/2\pi
\hbar c$ takes integers. In this case Eq. (37) becomes
\bb
\psi_0(r,\theta+2\pi)=\psi_0(r,\theta), \quad (\nu=0),
\ee     
which means $\psi_0$ is single valued. This is because the first
factor in Eq. (33) is also single valued in the present case. The
second case we are to consider is $\nu={1\over 2}$,
or $q\Phi/2\pi\hbar c$ takes half integers.
In this case Eq. (37) becomes
\bb
\psi_0(r,\theta+2\pi)=-\psi_0(r,\theta), \quad
(\nu=\textstyle{1\over2}).
\ee     
Though this is not convenient in polar coordinates, it may be easily
treated in parabolic coordinates.

Now we introduce the parabolic coordinates ($\xi,\eta$) whose
relation with ($x,y$) and ($r,\theta$) are given by
\bb
x=\textstyle{1\over2}(\xi^2-\eta^2),\quad y=\xi\eta,
\ee             
\bb
\xi=\sqrt{2r}\cos{\theta\over 2},\quad
\eta=\sqrt{2r}\sin{\theta\over 2}.
\ee              
In these coordinates, Eqs. (38) and (39) become
\bb
\psi_0(-\xi,-\eta)=\psi_0(\xi,\eta),\quad (\nu=0)
\ee                
and
\bb
\psi_0(-\xi,-\eta)=-\psi_0(\xi,\eta),\quad (\nu=\textstyle{1\over2})
\ee                
respectively, where for convenience we have used the same notation
$\psi_0$ to denote the wave function in parabolic coordinates. It is
easy to see that other values of $\nu$ in Eq. (37) renders
$\psi_0(\xi,\eta)$ multivalued and thus are difficult to deal with.
Though $\psi_0$ is double valued in polar coordinates in the case
$\nu={1\over 2}$, it becomes single valued in the parabolic
coordinates. This is essentially because a $\xi\eta$ plane covers
the $xy$ plane twice, which is obvious from the relation
$x+iy=(\xi+i\eta)^2/2$.

In the parabolic coordinates Eq. (34) becomes
\bb
(\partial_\xi^2+\partial_\eta^2)\psi_0+k^2(\xi^2+\eta^2)\psi_0
+4\beta k\psi_0=0,
\ee               
where
\bb
k={\sqrt{2\mu E}\over \hbar},\quad \beta={\mu\kappa\over\hbar^2 k}.
\ee              
Note that $E>0$ since we are considering scattering states, and
$\beta$ is dimensionless. Equation (44) can be solved by separation
of variables. Let
\bb
\psi_0(\xi,\eta)=v(\xi)w(\eta),
\ee                
we have for $v$ and $w$ the following equations:
\bb
v''+k^2\xi^2 v+\beta_1 k v=0,
\ee                      
\bb
w''+k^2\eta^2 w+\beta_2 k w=0,
\ee                      
where $\beta_1+\beta_2=4\beta$, and primes denote differentiation
with respect to argument. The general solution of Eq. (44) can be
obtained by superposition of solutions of the form (46) over the
parameter $\beta_1$. For the scattering problem at hand we will see,
however, that a single $\beta_1$ is sufficient. No superposition is
necessary. Specifically, we are looking for solutions that have the
asymptotic property
\bb
\psi_0\sim e^{ikx},\quad {\rm for}~~ x\to-\infty.
\ee                   
This represents  particles incident in the $+x$ direction, as is
easily verified by using Eq. (36).
In the parabolic coordinates it becomes
\bb
\psi_0\sim e^{ik(\xi^2-\eta^2)/2},\quad {\rm for}~~\eta\to\infty
~~{\rm and~all}~~\xi.
\ee                   
This can be satisfied only if
\bb
v(\xi)=e^{ik\xi^2/2}
\ee                 
and $w(\eta)$ has the asymptotic form
\bb
w(\eta)\sim e^{-ik\eta^2/2}, \quad {\rm for}~~ \eta\to\infty.
\ee          
It is easy to verify that $v(\xi)$ given by Eq. (51) does satisfy
Eq. (47) with
$\beta_1=-i$. Then the constant $\beta_2$ in Eq. (48) is given by
$\beta_2=4\beta+i$. The subsequent discussions depend on the value
of $\nu$, and we should distinguish between the two cases $\nu=0$
and $\nu={1\over2}$.

For $\nu=0$ we define a new function $u(\eta)$ by
\bb
w(\eta)=e^{-ik\eta^2/2}u(\eta),
\ee          
then Eq. (48) becomes
\bb
u''-2ik\eta u'+4\beta k u=0.
\ee         
On account of Eqs. (51) and (53), the condition (42) now simply means
that $u(\eta)$ is an even function of $\eta$:
\bb
u(-\eta)=u(\eta).
\ee           
It is easy to find the solution of Eq. (54) that satisfies this
condition:
\bb
u(\eta)=c_1 F(i\beta,\textstyle{1\over2}, ik\eta^2),
\ee           
where $c_1$ is a normalization constant. Collecting Eqs. (46), (51),
(53), and (56) we obtain the solution
\bb
\psi_0=c_1 e^{ik(\xi^2-\eta^2)/2}F(i\beta,\textstyle{1\over2},
ik\eta^2)=c_1 e^{ikx}F(i\beta,\textstyle{1\over2}, ik\eta^2).
\ee           
If in addition to $\nu=0$ we have $m_0=0$, i.e., for a pure Coulomb
potential, this is the required solution. Taking the limit $r\to
\infty$, and choosing the constant $c_1=e^{\beta\pi/2}\Gamma(1/2-i
\beta)/\sqrt\pi$, we have for $\psi_0$ the asymptotic form
\bb
\psi_0\to\exp[ikx-i\beta\ln k(r-x)]
+f_{\rm C}(\theta){\exp(ikr+i\beta\ln 2kr)\over \sqrt r},
\quad (r\to\infty)
\ee                  
up to the order $r^{-1/2}$, where
\bb
f_{\rm C}(\theta)={\Gamma(1/2-i\beta)\over\Gamma(i\beta)}
{\exp(i\beta\ln\sin^2\theta/2-i\pi/4)
\over\sqrt{2k\sin^2\theta/2}}.
\ee                 
The first term in the above equation represents the incident wave
while the second represents the scattered one. Both of them 
are distorted by a logarithmic term in the phase
due to the long-range nature of the Coulomb field.
Despite these distortions, it can be shown that the scattering cross
section is given by
\bb
\sigma(\theta)=|f_{\rm C}(\theta)|^2,
\ee           
where the subscript C indicates pure Coulomb scattering. 
Using the mathematical formulas
\bb
|\Gamma(\pm i\beta)|^2={\pi\over \beta\sinh \beta\pi},\quad
|\Gamma({\textstyle{1\over2}}\pm i\beta)|^2={\pi\over\cosh \beta\pi},
\ee              
we arrive at
\bb
\sigma_{\rm C}(\theta)={\beta\tanh \beta\pi\over 2k\sin^2\theta/2}.
\ee             
This is the result obtained in Ref. [25]. If $m_0\ne0$, i.e., if
$q\Phi/2\pi\hbar c$ takes nonzero integers, the solution (57) is in
problem, however. This is because $\psi_0({\bf r}=0)=c_1\ne0$, and
according to Eq. (33) $\psi({\bf r}=0)=c_1 e^{-im_0\theta}$, which is
not well defined since $\theta$ is not well defined at the origin.
The correct solution for $m_0\ne0$ should be
\bb
\psi_0=c_1[e^{ikx}F(i\beta,\textstyle{1\over2}, ik\eta^2)
-e^{ikr}F(\textstyle{1\over2}-i\beta,1, -2ikr)],
\ee           
where the second term in the square bracket also solves Eq. (34) with
the condition (38), and does not affect the boundary condition (49).
We have now $\psi_0({\bf r}=0)=0$ and no problem
arises. Due to this additional term, the solution now behaves at
infinity like
\bb
\psi_0\to \psi_{\rm in}+\psi_{\rm sc}+\psi_{\rm st},
\quad (r\to\infty)
\ee                  
where $\psi_{\rm in}$ and $\psi_{\rm sc}$ represent the incident and
scattered waves which are given by the first and second terms in Eq.
(58), respectively, and $\psi_{\rm st}$ represents a stationary wave
which comes from the second term in Eq. (63) and is given by
\bb
\psi_{\rm st}=-e^{i\delta_0}\sqrt{2\over \pi k}{\cos(kr+\beta\ln 2kr+
\delta_0-\pi/4)\over \sqrt r},
\ee             
where
\bb
\delta_0=\arg\Gamma(\textstyle{1\over2}-i\beta).
\ee              
Since the second term in Eq. (63) is in fact the $s$-wave term in the
partial wave expansion for a pure Coulomb  field, the logarithmic
distortion in its asymptotic form mentioned before becomes clear
here. Similar distortions appear in all partial waves regardless of
whether the AB potential is present. The first term in Eq. (64) gives
an incident current in the $+x$ direction (when $x\to-\infty$). The
second gives a scattered one in the radial direction (the component
in the $\theta$ direction can be ignored when $r\to\infty$) and leads
to the cross section  $\sigma_{\rm C}(\theta)$ obtained above. The
third term, as a stationary wave, contributes nothing to the cross
section. There are, however, interference terms. The interference of
the first term with the subsequent ones does not lead to physically
significant results. However, the interference of the second and the
third terms actually gives rise to an additional term in the cross
section, which will be denoted by $\sigma_\times(\theta)$.
The differential cross section in the present case is thus given by
\bb
\sigma_1(\theta)=\sigma_{\rm C}(\theta)+\sigma_\times(\theta),
\ee               
where
\bb
\sigma_\times(\theta)=-{\sqrt{\beta\tanh\beta\pi}\over\sqrt\pi k}
{\cos(\delta_0+\delta_1-\beta\ln\sin^2\theta/2)\over|\sin\theta/2|},
\ee              
and
\bb
\delta_1=\arg\Gamma(i\beta).
\ee             
In the neighbourhood of $\theta=0$, $\sigma_\times(\theta)$ oscillates
rapidly and thus the total contribution in a finite (but small)
interval of $\theta$ may be neglected. For large $\theta$, especially
near $\theta=\pi$, however, $\sigma_\times(\theta)$ gives considerable
contribution. It is remarkable that $\sigma_\times(\theta)$ is not
positive definite and thus $\sigma_1(\theta)$ may become negative
somewhere. This means that the particles move toward the origin at
some directions. To the best of our knowledge, similar results were
not encountered previously in the literature. Though the differential
cross section $\sigma_1(\theta)$ may become negative at some
direction, it does not cause any trouble physically because the total
cross section is positive (actually positively infinite due to the
long range nature of the potentials). Indeed, $\sigma_\times(\theta)$
gives a finite contribution (positive or negative) to the total cross
section, while $\sigma_{\rm C}(\theta)$ gives a positively infinite
one.

Now we turn to the case $\nu={1\over2}$. In this case we make the
transformation
\bb
w(\eta)=e^{-ik\eta^2/2}\eta u(\eta),
\ee          
then  the equation for $u$ reads
\bb
\eta u''+2(1-ik\eta^2)u'+2k(2\beta-i)\eta u=0.
\ee         
The condition (43) means that $u(\eta)$ is an even function of $\eta$.
The required solution can be found to be
\bb
u(\eta)=c_2 F(i\beta+\textstyle{1\over2},{3\over2}, ik\eta^2),
\ee           
where $c_2$ is a normalization constant. Collecting Eqs. (46), (51),
(70), and (72) we obtain the solution
\bb
\psi_0=c_2 e^{ikx}\eta
F(i\beta+\textstyle{1\over2},{3\over2},ik\eta^2).
\ee           
Here two remarks should be made. First, as a function of $r$ and
$\theta$, $\psi_0$ is double valued, so that $\psi$ is single valued
[cf. Eq. (33) where now $\nu=\frac12$] . Second, as a consequence of
Eq. (43) and obvious from the above result, we have
$\psi_0({\bf r}=0)=0$ here, so that $\psi$ is well defined at the
origin. We choose
$$
c_2= 2\sqrt{k \over \pi}\exp\left({\beta\pi\over2}-i{\pi\over4}\right)
\Gamma(1-i\beta),
$$
then the asymptotic form of $\psi_0$ is given by
\bb
\psi_0\to\exp[ikx-i\beta\ln k(r-x)]{\sin\theta/2\over|\sin\theta/2|}
+f(\theta){\exp(ikr+i\beta\ln 2kr)\over \sqrt r},
\quad (r\to\infty)
\ee                  
where
\bb
f(\theta)={\beta\Gamma(-i\beta)\over\Gamma(1/2
+i\beta)}{\exp(i\beta\ln\sin^2\theta/2+i3\pi/4)
\over\sqrt{2k}\sin^2\theta/2}.
\ee                 
Again note that both terms are double valued. The double valueness
does not cause much trouble in the calculation. Using the formulas
(61) the cross section can be shown to be
\bb
\sigma_2(\theta)=|f(\theta)|^2
={\beta\coth \beta\pi\over 2k\sin^2\theta/2}.
\ee             
This has the same angular distribution as $\sigma_{\rm C}(\theta)$,
but the dependence on other parameters is quite different.

If we ignore the relation $\kappa=Zq^2$ and treat $\kappa$ as an
independent parameter,  we may set $\kappa=0$ in the above results
(note that $Z=0$ is not allowed in our formalism). Then we have
\bb
\sigma_1(\theta)=0,\quad 
\sigma_2(\theta)={1\over 2\pi k\sin^2\theta/2}.
\ee             
These are the AB scattering cross sections for the corresponding
values of $\nu$.

Finally we point out that the cross sections (62), (67), and (76),
when expressed in terms of the classical velocity
$v_{\rm c}=\hbar k/\mu$ instead of $k$, involve $\hbar$ explicitly.
In the classical limit, $\hbar\to0$, $\beta=\kappa/\hbar v_{\rm c}
\to\infty$ (this is actually realized in the low energy limit),
we see that $\sigma_\times(\theta)$ is negligible in compared with
$\sigma_{\rm C}(\theta)$, and both $\tanh \beta\pi$ and
$\coth \beta\pi$ tend to $\pm 1$. So we have in this limit
\bb
\sigma_{\rm C}(\theta)=\sigma_1(\theta)=\sigma_2(\theta)
={|\kappa|\over 2\mu v_{\rm c}^2 \sin^2\theta/2},
\ee           
which is the classical scattering cross section for a pure Coulomb
field in two dimensions. This result implies that the AB potential
has no significant effect in the classical limit as expected.

\section{Summary and discussions}           

In this paper we propose an $n$-body Schr\"odinger equation for
particles carrying magnetic flux as well as electric charges.
The ratio of electric charge to magnetic flux is the same for all
particles. The two-body problem is studied in detail. The bound
state problem is exactly solved in the general case, while the
scattering problem is exactly solved in two special cases.

The original intention of this work is to describe the CS vortex
solitons by a simple quantum mechanical model. If the sizes of the
solitons are small, the AB potential may be a good approximation
in describing the charge-flux interaction. On the other hand, the
real charge-charge interaction may be quite complicated, thus the
Coulomb potential used here may be questionable.  If a better form
$V(q_a,q_b,|{\bf r}_a-{\bf r}_b|)$ can be found for the interaction
potential of charge $q_a$ at ${\bf r}_a$ and charge $q_b$ at
${\bf r}_b$, then the $n$-body equation may be improved by
substituting this potential for $q_aq_b/|{\bf r}_a-{\bf r}_b|$ in
Eq. (2b). In this case the last term $q_1q_2/r$ in the two-body
relative Hamiltonian (11b) should be replaced by $V(q_1, q_2,r)$.
With an improved potential, the Schr\"odinger equation might become
more difficult to solve, however. Therefore, the model studied in this
paper, even though it cannot well describe the interaction of the
vortex solitons, may have some interest in itself since it allows
exact analysis to some extent.

Several aspects of this model that need further studies may be:
the dynamical symmetry of the two-body system, the scattering problem
for general value of $\nu$, and finally, the relativistic
generalization of the model.
                                    
\section*{Acknowledgment}

The author is grateful to Professor Guang-jiong Ni for encouragement.
This work was supported by the
National Natural Science Foundation of China.



\begin{thebibliography}{99}

\bibitem{1}S. K. Paul and A. Khare, Phys. Lett. {\bf B174}, 420
            (1986); {\it ibid.} {\bf B182}, 415(E) (1986).
\bibitem{2}J. Hong, Y. Kim, and P. Y. Pac, Phys. Rev. Lett.
      {\bf 64}, 2230 (1990).
\bibitem{3}R. Jackiw and E. J. Weinberg, Phys. Rev. Lett. {\bf 64},
          2234 (1990).
\bibitem{4}R. Jackiw, K. Lee, and E. J. Weinberg,
           Phys Rev. D {\bf 42}, 3488 (1990).
\bibitem{5}R. Jackiw and S.-Y. Pi, Phys. Rev. Lett. {\bf 64},
           2969 (1990); Phys. Rev. D {\bf 42}, 3500 (1990).
\bibitem{6}Z. F. Ezawa, M. Hotta, and A. Iwazaki, Phys. Rev. Lett.
           {\bf 67}, 411 (1991); {\bf 67}, 1475(E) (1991).
\bibitem{7}R. Jackiw and S.-Y. Pi, Phys. Rev. Lett.
          {\bf 67}, 415 (1991);
           Phys. Rev. D {\bf 44}, 2524 (1991).
\bibitem{8}Q.-G. Lin, Phys. Rev. D {\bf 48}, 1852 (1993).
\bibitem{9}H. B. Nielsen and P. Olesen, Nucl. Phys. {\bf B61}, 45
          (1973).
\bibitem{10}Y. Aharonov and D. Bohm, Phys. Rev. {\bf 115}, 485 (1959).
\bibitem{11}S. N. M. Ruijsenaars, Ann. Phys. (NY) {\bf 146}, 1 (1983);
        R. Jackiw, Ann. Phys. (NY) {\bf 201}, 83 (1990).
\bibitem{12}C. Manuel and R. Tarrach, Phys. Lett. {\bf B268}, 222
    (1991); {\it ibid.} {\bf B328}, 113 (1994); O. Bergman and
    G. Lozano, Ann. Phys. (NY) {\bf 229}, 416 (1994);
    G. Amelino-Camelia and D. Bak, Phys. Lett. {\bf B343}, 231 (1995).

\bibitem{13}S.P. Alliluev, Sov. Phys. JETP {\bf 6}, 156 (1958).
\bibitem{14}T.-I. Shibuya and C.E. Wulfman, Am. J. Phys. {\bf 33},
            570 (1965).
\bibitem{15}B. Zaslow and M.E. Zandler, Am. J. Phys. {\bf 35},
             1118 (1967).
\bibitem{16}J.-Y. Zeng and W.-S. Wan, College Physics 7(3), 1 (1988)
            (in Chinese);
           X.L. Yang, M. Lieber, and F.T. Chan, Am. J. Phys. {\bf 59},
            231 (1991).
\bibitem{17}A. Cisneros and H.V. McIntosh, J. Math. Phys. {\bf 10},
           277 (1969).
\bibitem{18}D.S. Bateman, C. Boyd, and B. Dutta-Roy, Am. J. Phys.
            {\bf 60}, 833 (1992).
\bibitem{19}A. Guha and S. Mukherjee, J. Math. Phys. {\bf 28}, 840
               (1987).
\bibitem{20}M. Kibler and T. Negadi, Phys. Lett. {\bf A124}, 42
             (1987).
\bibitem{21}G. E. Draganescu, C. Campiogotto, and M. Kibler,
           Phys. Lett. {\bf A170}, 339 (1992).
\bibitem{22}V. M. Villalba, Phys. Lett. {\bf A193}, 218 (1994).
\bibitem{23}L. Chetonani, L. Guechi, and T. F. Harman,
           J. Math. Phys. {\bf 30}, 655 (1989).
\bibitem{24}R. Dutt, A. Gangopadhyaya, and U. P. Sukhatme,
           Am. J. Phys. {\bf 65}, 400 (1997).
\bibitem{25}G. Barton, Am. J. Phys. {\bf 51}, 420 (1983); Q.-G. Lin,
   {\it ibid.} {\bf 65}, 1007 (1997); M. J. Moritz and H. Friedrich,
   {\it ibid.} {\bf 66}, 274 (1998).
\end{thebibliography}
\end{document}